\documentclass{raa}          

\usepackage{graphicx,times}             % for PS/EPS graphics inclusion
\usepackage{natbib}
\usepackage{amssymb,amsmath}
\usepackage[T1]{fontenc}
\usepackage{longtable}
\usepackage{newtxtext,newtxmath}
\usepackage{placeins}
\usepackage{threeparttable}
\usepackage{rotating}
\usepackage{booktabs}
\usepackage[utf8]{inputenc}
\usepackage{multirow} 
\usepackage{aliascnt}
\usepackage{enumitem}
\usepackage{xcolor}
\usepackage{gensymb}
\usepackage{tabularx}
\usepackage{orcidlink}
\usepackage{fontawesome}
\bibpunct{(}{)}{;}{a}{}{,}
\hypersetup{colorlinks=true,allcolors=blue}

\begin{document}
%   \subtitle{I. Place Your Subtitle Here}
\title{Dust masses of Upper Scorpius disks and their statistics}
%   \subtitle{I. Place Your Subtitle Here}
   \volnopage{Vol.0 (20xx) No.0, 000--000}      %%preserved for Editor. DOn't remove!
   \setcounter{page}{1}          %%starting page, preserved for Editor. DOn't remove!

\author{Huiyi Deng
      \inst{1}
   \and Yao Liu
      \inst{1}
   \and Min Fang
      \inst{2}	  
   }

\institute{School of Physical Science and Technology, Southwest Jiaotong University, Chengdu 610031, China; {\it yliu@swjtu.edu.cn}\\
  \and
  Purple Mountain Observatory, Chinese Academy of Sciences, 10 Yuanhua Road, Nanjing 210023, China \\ 
\vs\no
   {\small Received 2026 month day; accepted 2026 month day}}

\abstract{The total dust mass of protoplanetary disks is a key property that determines the potential for planet formation. Upper Scorpius (USco), with an age of ${\sim}\,5\,{-}\,12\,{\rm Myr}$, provides an important laboratory for investigating the evolution of dust reservoirs on timescales comparable to those of planet formation. In this work, we analyze the dust mass distribution of 136 full and transitional disks in USco using the largest ALMA 0.88\,mm continuum sample currently available for the region. To improve the accuracy of dust mass estimates, we construct a grid of 918 self-consistent radiative transfer models that account for stellar-mass-dependent disk sizes, accretion rates, and porous dust properties. The models yield a new calibration between dust temperature and stellar luminosity, predicting systematically higher dust temperatures for low-mass stars than commonly adopted prescriptions. Using these revised dust temperatures together with porous dust opacities, we derive dust masses for the USco disks and compare them with those of a younger sample in the Chamaeleon~I star-forming region. We find a median dust mass of $0.95\,M_{\oplus}$ for the USco disks, approximately six times lower than that of the Chamaeleon~I disks ($5.69\,M_{\oplus}$), providing strong evidence for substantial depletion of millimeter-sized dust grains over the first several Myr of disk evolution. We confirm the previous finding of a highly significant correlation between stellar mass and dust mass, with a slightly steeper relation in USco than in Chamaeleon~I. The low dust masses observed in USco, together with their comparison to mature exoplanetary systems, suggest that a large fraction of the primordial solid reservoir has already been incorporated into larger bodies, removed by radial drift, or hidden from millimeter observations by ages of ${\sim}\,5\,{-}\,12\,{\rm Myr}$. 
}

\authorrunning{Deng et al.}            %author_head in even pages
\titlerunning{Dust masses of Upper Scorpius disks}  % title_head in odd pages
\maketitle

\section{Introduction}           %% first-level sections will be auto-capitalized
\label{sect:intro}

Protoplanetary disks composed of gas and dust are the birthplace of planets \citep[e.g.,][]{Williams2011,Raymond2022,Drazkowska2023}. As the initial building blocks of planets, submicron-sized dust grains in protoplanetary disks collide and coagulate into larger aggregates, eventually forming pebbles, planetesimals, and planetary embryos  \citep[]{Lissauer1993,Dullemond2005,Birnstiel2016}. Dust grains also strongly influence the thermal and chemical structure of disks by regulating the opacity and mediating radiative heating and cooling processes \citep[e.g.,][]{Dullemond2007,Henning2013}. In addition, dust particles provide the surface area on which chemical reactions can occur to potentially form complex organic molecules \citep{Garrod2006,Oberg2023}. Therefore, constraining the dust content in protoplanetary disks is essential not only for understanding the efficiency and timescale of planet formation, but also for providing key observational constraints on models of grain growth, radial drift, and disk chemistry \citep{Birnstiel2024}.

The total amount of dust material in protoplanetary disks is one of the most important properties that characterize the potential for planet formation. (Sub-)millimeter continuum observations is the primary method for estimating dust masses \citep[e.g.,][]{Andrews2020,Miotello2023}. Under the assumption that the continuum emission is optically thin, the dust mass can be derived analytically from the observed flux density ($F_{\nu}$) with  
\begin{equation}
M_{\rm dust} = \frac{F_{\nu}D^2}{\kappa_{\nu}B_{\nu}(T_{\rm dust})},
\label{eqn:mana}
\end{equation}
where $B_{\nu}(T_{\rm dust})$ refers to the Planck function given at the observed frequency $\nu$ and dust temperature $T_{\rm dust}$, the distance to the object is denoted as $D$, and $\kappa_{\nu}$ stands for the dust absorption opacity. The advent of the Atacama Large Millimeter/submillimeter Array (ALMA) has revolutionized this field by sensitive continuum surveys of nearby star-forming regions, for instance Taurus \citep{Long2019}, Lupus \citep{Ansdell2016}, Chamaeleon~I \citep{Pascucci2016}, Ophiuchus \citep{Williams2019}, and Upper Scorpius \citep{Barenfeld2016}. These surveys have revealed a systematic decline in dust masses with the age of star-forming regions, indicating substantial dust evolution over Myr timescales \citep[e.g.,][]{Andrews2013,Pascucci2016,Grant2021,Manara2023}. However, significant uncertainties remain in dust mass estimates, particularly in terms of the adopted dust temperature and dust opacity \citep{Beckwith1991,Miotello2023,Liuy2024}. Refining these assumptions is therefore important for obtaining more accurate constraints on the solid mass budget available for planet formation.

Upper Scorpius association (hereafter USco) is a benchmark region for investigating the evolution of protoplanetary disks. As part of the Scorpius-Centaurus OB association, USco has an estimated age of $5\,{-}\,12\,{\rm Myr}$ \citep{Pecaut2012,David2019,Luhman2020}, which is substantially older than other nearby star-forming regions such as Taurus \citep[$1\,{-}\,3\,{\rm Myr}$,][]{Krolikowski2021} and Chamaeleon~I \citep[$1\,{-}\,2\,{\rm Myr}$,][]{Galli2021}. At this evolutionary stage, disks are expected to have undergone significant dust evolution and dispersal \citep{Williams2011,Testi2014}. Consequently, USco provides an ideal laboratory for studying the evolution of dust reservoirs on timescales comparable to those of planet formation. The disk population in USco has been extensively surveyed at (sub-)millimeter wavelengths. \citet{Carpenter2014} first presented ALMA 0.88\,mm continuum observations of 20 K- and M-type stars hosting protoplanetary disks. The sample was subsequently expanded  to 106 disks by \citet{Barenfeld2016}. More recently, \citet{Carpenter2025} carried out a new ALMA 0.88\,mm continuum survey and increased the sample to 202 systems, providing the most comprehensive census of the USco disk population to date.

In this work, based on the expanded sample compiled by \citet{Carpenter2025}, we analyze the dust masses of USco disks by incorporating updated dust temperatures derived from physically motivated disk models and porous dust opacities. We investigate the statistical properties of the dust mass distribution, and discuss their implications for dust evolution and planet formation. Sect.~\ref{sec:sample} describes the sample selection and the derivation of the host stellar properties. In Sect.~\ref{sec:tdust}, we construct a grid of self-consistent radiative transfer models that account for stellar-mass-dependent disk size and accretion rate, with the goal of calibrating the relation between stellar luminosity and dust temperature over a broad range of stellar types. Sect.~\ref{sec:mdust} presents the inferred dust masses and their statistical properties. Finally, Section~\ref{sec:summary} summarizes our main results.

\begin{figure}
\centering
\includegraphics[width=0.9\textwidth]{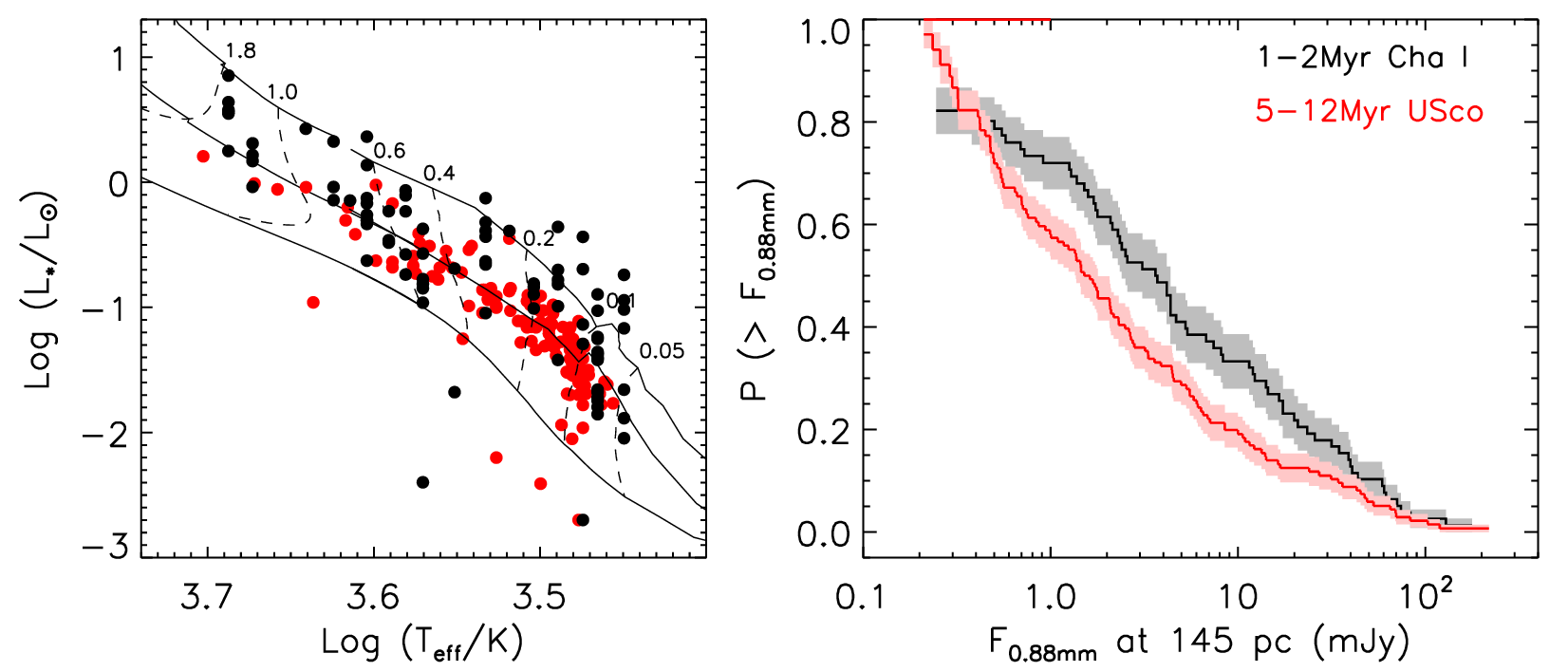}
\caption{Properties of the sample investigated in this work. {\it Left panel: } Hertzsprung–Russell diagram, in which the solid lines represent isochrones of 0.5, 3.0 and 20\,Myr from the \citet{Baraffe2015} models for $T_{\rm eff}\,{\lesssim}\,3900\,{\rm K}$ and  from the \citet{Feiden2016} models for hotter stars. The evolutionary tracks for different stellar masses are indicated with the dashed curves. {\it Right panel: } cumulative distribution function of the 0.88\,mm fluxes scaled at a distance of 145\,pc. In both panels, symbols with red color refer to the old USco disks, while the young Chamaeleon~I sample is shown with black symbols.}
\label{fig:sample}
\end{figure}

\section{The USco disks and their comparative sample}
\label{sec:sample}

Using sensitive ALMA 0.88\,mm continuum observations, \citet{Carpenter2025} compiled a sample of 202 protoplanetary disks in USco. In this work, we focus on the 136 sources classified as either ``full'' or ``transitional'' disks based on their infrared color excesses, see Table~1 of \citet{Carpenter2025}. The sample spans spectral types from K1 to M5.5. Stellar effective temperature ($T_{\rm eff}$) and luminosity ($L_{\star}$) were adopted from \citet{Fang2023} and \citet{Fang2025}. The effective temperatures were derived by converting spectral types using the calibration of \citet{Fang2017}, while the stellar luminosities were estimated either from the extinction-corrected absolute flux density at $7500\textup{~\AA}$ or from the spectral energy distribution fitting in the optical wavelength range. 

The red symbols in the left panel of Figure~\ref{fig:sample} show the USco sample in the Hertzsprung–Russell diagram. Using the derived $T_{\rm eff}$ and $L_{\star}$, we estimated the stellar mass ($M_{\star}$) by interpolating pre-main-sequence evolutionary tracks. For objects with $T_{\rm eff}\,{\lesssim}\,3900\,{\rm K}$ (M dwarfs), we adopted the models of \citet{Baraffe2015}, while for hotter stars (spectral types K and earlier) we used the non-magnetic tracks of \citet{Feiden2016}. This approach is motivated by the fact that the two sets of models show good agreement around $T_{\rm eff}\,{\sim}\,3900\,{\rm K}$. To investigate the evolution of dust masses with age, we selected the Chamaeleon~I star-forming region (hereafter Cha~I) as a comparison sample. \citet{Pascucci2016} presented ALMA 0.88\,mm continuum measurements for 93 Class~II or transitional disks in Cha~I, while \citet{Long2018} obtained deeper ALMA observations of 14 of these systems, improving the sensitivity by approximately a factor of five. From this sample, we selected the 78 disks with spectral types between K1 and M5.5 to provide a stellar-mass range comparable to that of the USco sample. The stellar properties of the Cha~I sources were adopted from \citet{Manara2023}, who derived them using the same methodology and evolutionary models employed in this work. The Cha I sources are shown with black symbols in Figure~\ref{fig:sample}. 

To characterize the distribution of 0.88\,mm continuum flux densities ($F_{\rm 0.88mm}$), we constructed cumulative distributions using the Kaplan-Meier product-limit estimator implemented in the \texttt{ASURV} package, which properly accounts for censored data and upper limits \citep{Feigelson1985,Isobe1986}. Prior to the survival analysis, the observed flux densities were scaled to a common distance of 145\,pc using distances inferred from Gaia parallaxes \citep{Gaia2023}. For sources without Gaia distance measurements, we adopted distances of 145\,pc and 190\,pc for the USco and Cha~I disks, respectively. The resulting cumulative distributions are shown in the right panel of Figure~\ref{fig:sample}. The median 0.88\,mm flux density of the USco disks is 1.5\,mJy, a factor of 2.4 lower than that of the younger Cha~I disks (3.6\,mJy), indicating substantial evolution of the solid component over the ${\sim}\,5\,{-}\,10\,{\rm Myr}$ age interval between the two regions.

\section{Dust temperature as a function of stellar luminosity}
\label{sec:tdust}

As given in Eq.~\ref{eqn:mana}, one needs to know the dust temperature ($T_{\rm dust}$) to calculate the dust mass. A value of 20\,K is a common choice in literature studies \citep[e.g.,][]{Pascucci2016,Manara2023}. However, as the stellar irradiation is the main heating source for protoplanetary disks \citep{Chiang1997,Dullemond2007}, especially in the Class II evolutionary stage, dust temperature should in principle vary with stellar parameters. Based on a set of radiative transfer models, \citet{Andrews2013} obtained a relation between stellar luminosity and dust temperature $T_{\rm dust}\,{=}\,25\,(L_{\star}/L_{\odot})^{0.25}\,\rm{K}$. The scaling of $T_{\rm dust}$ with $L_{\star}$ was further investigated for disks around low-mass stars and brown dwarfs \citep[e.g.,][]{Daemgen2016,vanderPlas2016,Hendler2017}, and the relation, $T_{\rm dust}\,{=}\,22\,(L_{\star}/L_{\odot})^{0.16}\,\rm{K}$, appears flatter than that suggested by \citet{Andrews2013}.  

The studies discussed above have substantially improved our understanding of the dependence of dust temperature on stellar luminosity. Nevertheless, several important issues remain. First, most previous investigations assume a fixed disk outer radius (typically 100\,AU) across the entire range of stellar masses. However, spatially resolved ALMA observations show that disks around lower-mass stars are systematically smaller than those around higher-mass stars \citep[e.g.,][]{Tripathi2017,Andrews2018b,Andrews2020}. Because disk size determines the characteristic distance of dust grains from the central heating source, it has a direct impact on the dust temperature \citep{vanderPlas2016,Hendler2017}. Second, viscous heating is often neglected in the radiative transfer modeling. Observations indicate that the disk accretion rate generally decreases with stellar mass \citep[e.g.,][]{Fang2009,Manara2023}. Although low-mass stars and brown dwarfs generally exhibit lower accretion rates, the contribution of accretion luminosity to the disk energy budget may still be non-negligible because their stellar luminosities are also substantially lower. Third, recent infrared and millimeter observations suggest that dust grains in protoplanetary disks might be highly porous \citep[e.g.,][]{Ginski2023,Tazaki2023,Zhang2023}. Porous dust grains differ a lot from compact grains in terms of scattering and absorption behaviors, which in turn have a direct impact on the dust temperature and the inferred dust mass \citep{Kirchschlager2019,Brunngraber2021,Liuy2024}.  

In this section, we recalibrate the relation between stellar luminosity and dust temperature by generating a grid of radiative transfer models computed with the \texttt{HOCHUNK3D} code \citep{Whitney2013}. The model grid spans a broad range of stellar masses, from intermediate-mass stars to planetary-mass objects. Unlike previous studies, our models incorporate stellar-mass-dependent disk sizes and accretion rates, allowing us to assess their impact on the dust temperature over a wide range of stellar luminosities. 

\subsection{Dust density distribution}
We considered a flared disk with an inner radius ($R_{\rm in}$) and outer boundary ($R_{\rm out}$). The disk inner radius is fixed to the dust sublimation radius $0.07\left(L_{\star} / L_{\odot}\right)^{0.5}{\rm AU}$ \citep{Dullemond2001}. The model contains two dust grain populations, a small grain population (SGP) and a large grain population (LGP). The idea of introducing two grain populations is stimulated to mimic the effect of dust settling in a simple way \citep{Andrews2011}. The scale height of the SGP follows a power law of
\begin{equation}
h(R) = h_{100}\times\left(\frac{R}{100\,\rm{AU}}\right)^\Psi.
\label{eq:heightgas}
\end{equation}
The flaring index is denoted with $\Psi$, while $h_{100}$ refers to the scale height at a radial distance of $R\,{=}\,100\,\rm{AU}$. On the contrary, the LGP is concentrated close to the midplane with a scale height of $\Lambda\,h$. The degree of dust settling is characterized by setting $\Lambda\,{=}\,0.2$. The SGP occupies a small fraction of the total dust mass, $(1-f)\,M_{\rm dust}$, while the LGP dominates the dust mass $f\,M_{\rm dust}$. In our models, we adopted $f\,{=}\,0.85$, which is consistent with multiwavelength modeling of protoplanetary disks \citep[e.g.,][]{vanderMarel2018,Schwarz2021,Zhang2021}. The surface density follows a power law with an exponential taper 
\begin{equation}
\Sigma(R)\,{=}\,\Sigma_{\rm c}\left(\frac{R}{R_{\rm c}}\right)^{-\gamma}{\rm exp}\left[-\left(\frac{R}{R_{\rm c}}\right)^{2-\gamma}\right],
\label{eqn:sigma}
\end{equation}
where the proportionality constant $\Sigma_{\rm c}$ is given by normalizing the total dust mass $M_{\rm dust}$, and the characteristic radius is given by $R_{\rm c}$. We fixed the gradient parameter $\gamma\,{=}\,1$ to reduce the degree of freedom. The adopted exponential surface density profile formally extends to infinity and therefore requires a numerical truncation. We set the disk outer radius $R_{\rm out}\,{=}\,5\,R_{\rm c}$ at which the surface density has already dropped by a factor of $e^{-5}\,{\approx}\,0.7\%$, such that nearly all of the disk mass is enclosed. This choice also yields disk outer sizes broadly consistent with the measured gas disk radii that typically extend several times beyond the characteristic dust radius \citep[e.g.,][]{Ansdell2018,Long2022,RuizRodriguez2025}. The dust volume density is given by
\begin{equation}
\rho_{\rm{SGP}}(R,z)\,{=}\,\frac{(1{-}f)\Sigma(R)}{\sqrt{2\pi}h}\,\exp\left[-\frac{1}{2}\left(\frac{z}{h}\right)^2\right],\\
\label{eqn:sgp}
\end{equation}
\begin{equation}
\rho_{\rm{LGP}}(R,z)\,{=}\,\frac{f\Sigma(R)}{\sqrt{2\pi}\Lambda h}\,\exp\left[-\frac{1}{2}\left(\frac{z}{\Lambda h}\right)^2\right].\\
\label{eqn:lgp}
\end{equation}

\subsection{Dust properties}
\label{sec:dustopac}

The dust composition was adopted from the DSHARP model \citep{Birnstiel2018}, which is composed of water ice \citep{Warren2008}, astronomical silicates \citep{Draine2003}, troilite \citep{Henning1996}, and refractory organic material \citep{Henning1996}, with volume fractions being 36\%, 17\%, 3\%, and 44\%, respectively. To introduce porosity, we mixed the optical constants of the DSHARP dust model with vacuum using the Bruggeman mixing rule \citep{Bruggeman1935}. The mixed refractive indices were used to calculate dust absorption and scattering properties with the Mie theory and \texttt{OpTool} \citep{Dominik2021}. The volume fraction of vacuum is controlled by the porosity $\mathcal{P}$. We chose $\mathcal{P}\,{=}\,0.8$ that is consistent with observational constraints of protoplanetary disks \citep[e.g.,][]{Zhang2023} and the cometary nucleus 67P/Churyumov–Gerasimenko \citep{Kofman2015,Jorda2016}. The bulk density of the porous dust grains is $\rho_{\rm s}\,{=}\,0.34\,\rm{g/cm^{3}}$. We note that compact dust particles can be considered as a limiting case of $\mathcal{P}\,{=}\,0$.

We assumed that the grain size distribution follows the power law ${\rm d}n(a)\,{\propto}\,{a}^{-3.5} {\rm d}a$ with a minimum grain size fixed to $a_{\rm{min}}\,{=}\,0.01\,\mu{\rm m}$. For the SGP, the maximum grain size was set to $a_{\rm{max}}\,{=}\,1\,\mu\rm{m}$. For the LGP, we adopted $a_{\rm{max}}\,{=}\,1\,\rm{mm}$ accounting for the effect of grain growth that is commonly seen in protoplanetary disks \citep[e.g.,][]{Liuy2017,Li2023,Shi2026}. By taking $a_{\rm{min}}\,{=}\,0.01\,\mu{\rm m}$ and $a_{\rm{max}}\,{=}\,1\,\rm{mm}$ in the calculation, porous grains with $\mathcal{P}\,{=}\,0.8$ feature an absorption coefficient at $\lambda\,{=}\,0.88\,{\rm mm}$ of $\kappa_{\rm 0.88mm}\,{=}\,0.84\,\rm{cm^2/g}$, which is about 4 times lower than compact grains with $\kappa_{\rm 0.88mm}\,{=}\,3.4\,\rm{cm^2/g}$.

\subsection{Heating mechanism}
\label{sec:heating}

In this work, both stellar irradiation and viscous heating are considered as heating sources in protoplanetary disks. For each model, the input stellar spectrum is simply represented by a blackbody with the corresponding effective temperature and luminosity. Moreover, we included accretion energy into the disk according to the $\alpha-$disk prescription \citep[e.g.,][]{Shakura1973,Kenyon1987}. Following \citet{Whitney2003}, the dissipated energy per unit volume generated by viscosity is assumed to be 
\begin{equation}
\frac{d\dot{E}_{\rm acc}}{dV} = \frac{3 G M_{\star} \dot{M}_{\rm acc}}
{\sqrt{32\pi^3}\,R^3 h(R)}\left(1-\sqrt{\frac{R_{\star}}{R}}\right)
\exp\left\{-\frac{1}{2}\left[\frac{z}{h(R)}\right]^2\right\},	
\end{equation}
where $\dot{M}_{\rm acc}$ is the disk accretion rate, $G$ is the gravitational constant, and $R_{\star}$ refers to the stellar radius.

Comparing with models heated purely by stellar irradiation, the inclusion of viscous heating significantly slows down the simulation. This is because photon packages originating from the dissipated energy start their journey right in the middle of the most optically thick regions (e.g., close to the midplane), which might require gazillions of absorption/re-emission events for them to escape the model space. In the simulation, we set the number of photons to $3\,{\times}\,10^7$. Roughly, it takes two to three days 
to run one model on a single CPU, highly depending on the optical depth of the disk. 

\subsection{Parameter space of the model grid}

Besides the fixed parameters described above, one needs to assign values of $L_{\star}$, $T_{\rm eff}$, $M_{\star}$, $R_{\star}$, $R_{\rm c}$, $h_{100}$, $\Psi$, $M_{\rm dust}$ and $\dot{M}_{\rm acc}$ for each model. We first sampled 34 values of $L_{\star}$ that are logarithmically distributed in the range of [$10^{-3}\,L_{\odot}$, $100\,L_{\odot}$]. Then, we interpolated the 2.5\,Myr isochrone of pre-main-sequence evolutionary tracks to obtain $T_{\rm eff}$, $M_{\star}$ and $R_{\star}$. We adopted the 2.5 Myr isochrone only to provide a self-consistent mapping between $L_{\star}$, $T_{\rm eff}$, $M_{\star}$ and $R_{\star}$ for constructing the grid of radiative transfer models. Since stellar luminosity is treated as an independent variable in our models, the resulting dust temperatures depend primarily on $L_{\star}$. Adopting older (e.g., 5\,$-$\,12\,Myr) isochrones changes the inferred $M_{\star}$ and $R_{\star}$ only modestly, and therefore has a negligible impact on the derived $T_{\rm dust}\,{-}\,L_{\star}$ relation. For $L_{\star}\,{<}\,0.4\,L_{\odot}$ (at approximately $T_{\rm eff}\,{<}\,3900\,{\rm K}$), we adopted the models presented by \citet{Baraffe2015}. We chose the nonmagnetic evolutionary models calculated by \citet{Feiden2016} for $L_{\star}\,{\geq}\,0.4\,L_{\odot}$. In the grid, there are three points for $h_{100}$: 5\,AU, 10\,AU, and 15\,AU. We considered three values for $\Psi$: 1.05, 1.15, and 1.25. We assumed three disk-to-stellar mass ratios, namely, 0.001, 0.01, and 0.1. Then, $M_{\rm dust}$ for each model is determined by assuming a dust-to-gas mass ratio of 0.01.   

\begin{figure}
\centering
\includegraphics[width=0.7\textwidth]{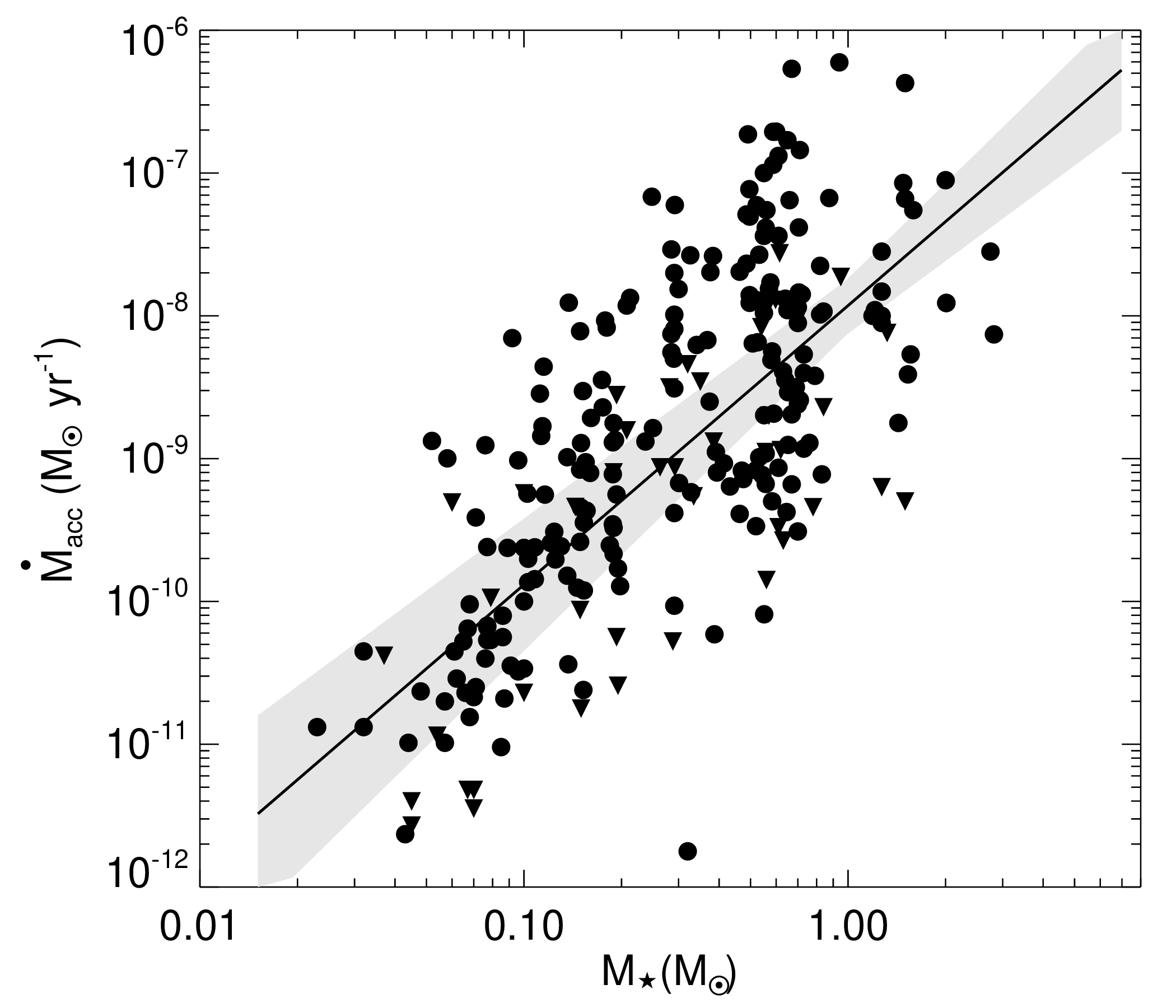}
\caption{Correlation between stellar mass and accretion rate. Data points were collected for disks in the Taurus, Chamaeleon, Lupus and Ophiuchus star-forming regions from \citet{Manara2023}. Measurements of the accretion rate are indicated with dots, whereas triangles represent upper limits. The solid line, expressed as 
${\rm log}{\dot M}_{\rm acc}\,{=}\,{-}7.93\,{+}\,1.95\,{\rm log}M_{\star}$, is the fitted relation derived from a Bayesian linear regression analysis that accounts for censored data. The gray shaded region marks the 95\% confidence interval on the relation.}
\label{fig:mstarmacc}
\end{figure}

The disk outer radius $R_{\rm out}$, which is linked with the characteristic radius via $R_{\rm out}\,{=}\,5\,R_{\rm c}$, is a crucial parameter affecting the temperature structure. ALMA observations reveal that $R_{\rm out}$ is stellar-mass dependent \citep{Tripathi2017,Andrews2020}, with smaller disks toward lower mass stars. By analyzing millimeter continuum images, \citet{Andrews2018b} determined the effective disk radius $R_{\rm eff}$ (i.e., the radius enclosing 68\% of the total flux) for 105 nearby protoplanetary disks. They obtained a linear correlation between the effective disk radius and stellar mass ${\rm log }\,R_{\rm eff}\,{=}\,1.77\,{+}\,0.58\,{\rm log}\,M_{\star}$. As demonstrated in \citet{Liuy2024}, the characteristic radius is a good representative for the effective disk radius. Therefore, we directly sampled $R_{\rm c}$ using the relation provided by \citet{Andrews2018b}. For a typical T Tauri star with $M_{\star}\,{=}\,0.5\,M_{\odot}$, the characteristic radius is $R_{\rm c}\,{\sim}\,40\,{\rm AU}$. 

The disk accretion rate directly determines the total energy release due to viscous heating. Many observations have shown that it varies across the stellar mass regime, with higher accretion rates toward more massive stars. \citet{Manara2023} complied a large sample of young stellar objects with measurements of accretion rate in nearby star-forming regions. Data points for 256 disks in the Taurus, Lupus, $\rho$\,Ophiuchus and Chamaeleon star-forming regions are shown in Figure~\ref{fig:mstarmacc}. We performed a correlation test by Cox proportional hazards model implemented in the \texttt{ASURV} package, properly accounting for upper limits \citep{Feigelson1985, Isobe1986}. The results show that the correlation between $M_{\star}$ and $\dot{M}_{\rm acc}$ is significant at a confidence level $P\,{<}\,0.01\%$. Then, we utilized the IDL routine \texttt{Linmix\_err} to conduct a Bayesian linear regression analysis that takes the errors and upper limits of measurements into account \citep{Kelly2007}. The best-fit relation, expressed as ${\rm log}{\dot M}_{\rm acc}\,{=}\,{-}7.93\,{+}\,1.95\,{\rm log}M_{\star}$, is indicated with the solid line in Figure~\ref{fig:mstarmacc}. Using this relation, we derived the accretion rate for each model. For a typical T Tauri star with $M_{\star}\,{=}\,0.5\,M_{\odot}$, the accretion rate is ${\dot M}_{\rm acc}\,{=}\,3\,{\times}\,10^{-9}\,M_{\odot}\,{\rm yr^{-1}}$.

To summarize, we sampled 34 different stellar luminosities ($L_{\star}$), corresponding to 34 stellar masses ($M_{\star}$). For each of the stellar luminosity bins, we considered three values for the scale height ($h_{100}$), three values for the disk flaring parameter ($\Psi$), and three values for the dust mass ($M_{\rm dust}$), which results in 27 different combinations. In total, our grid consists of $34\,{\times}\,27\,{=}\,918$ models.

\subsection{Results}

\begin{figure}
\centering
\includegraphics[width=0.7\textwidth]{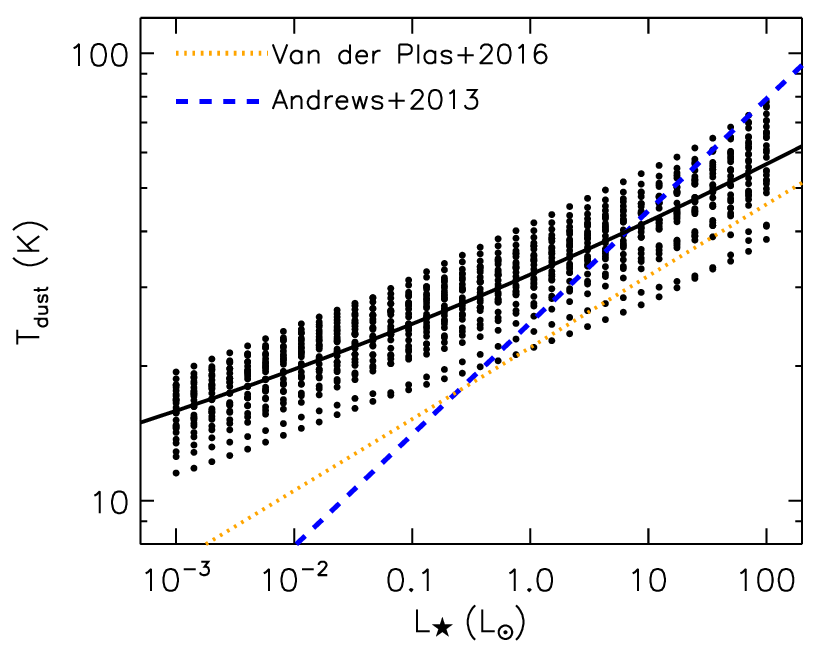}
\caption{Mass-averaged dust temperature as a function of stellar luminosity. The solid curve ${\rm log }\,T_{\rm dust}\,{=}\,1.506\,{+}\,0.115\,{\rm log}\,L_{\star}\,{+}\,0.004\,({\rm log}\,L_{\star})^2$ is the best-fit relation to our models. The blue dashed line refers to the relation $T_{\rm dust}\,{=}\,25\,(L_{\star}/L_{\odot})^{0.25}$ presented by \citet{Andrews2013}. The orange dotted line shows the relation $T_{\rm dust}\,{=}\,22\,(L_{\star}/L_{\odot})^{0.16}$ suggested by \citet{vanderPlas2016}.}
\label{fig:lstartdust}
\end{figure}

As our models are two dimensional, we divided the disk into 200 radial bins and 197 vertical bins to solve the radiative transfer problem, resulting in a total of $N_{\rm cells}\,{=}\,39,400$ cells. For each model, we calculated the mass-averaged dust temperature via 
\begin{equation}
 T_{\rm dust}\,{=}\,\frac{1}{M_{\rm dust}} \sum_{i=1}^{N_{\rm cells}}\sum_{j=1}^{N=2}m_{\rm dust}(i,j)\,t_{\rm dust}(i,j),
\end{equation}
where $m_{\rm dust}(i,j)$ and $t_{\rm dust}(i,j)$ are the dust mass and temperature in each cell $i$ for grain population $j$, respectively. Figure~\ref{fig:lstartdust} shows the mass-averaged dust temperature as a function of stellar luminosity. Although a dispersion in $T_{\rm dust}$ is present within each stellar luminosity bin, a clear trend of cooler disks around lower-mass stars is observed. We fit a second-order polynomial to the correlation 
between $T_{\rm dust}$ and $L_{\star}$. The best-fit relation is 
\begin{equation}
{\rm log }\,T_{\rm dust}\,{=}\,1.506\,{+}\,0.115\,{\rm log}\,L_{\star}\,{+}\,0.004\,({\rm log}\,L_{\star})^2.
\label{eqn:tdust}
\end{equation}
The scatter in $T_{\rm dust}$ within each stellar luminosity bin reflects the range of disk structural parameters explored in the grid. To illustrate the origin of this scatter, we examined the 27 models with $L_{\star}\,{=}\,0.1\,L_{\odot}$. For each value of the flaring index (or scale height), we calculated the mean dust temperature by averaging over the nine models spanning all combinations of the other two parameters. We find that increasing the flaring index from 1.05 to 1.25 raises the mean dust temperature from approximately 22\,K to 27\,K, while increasing the scale height from 5 AU to 15 AU raises it from approximately 22\,K to 28\,K. These results indicate that the scatter is primarily driven by variations in the disk geometry. More flared and geometrically thicker disks intercept a larger fraction of the stellar irradiation, resulting in higher dust temperatures. By contrast, the dependence of  $T_{\rm dust}$ on the adopted dust mass is comparatively weak over the parameter range explored in this work.

For $L_{\star}\,{\gtrsim}\,1\,L_{\odot}$, our relation predicts dust temperatures similar to those reported by \citet{Andrews2013}, although with a shallower slope. The discrepancies between our results and those of \citet{Andrews2013} and \citet{vanderPlas2016} become increasingly significant toward the low-stellar-mass regime. This difference arises because our models incorporate stellar-mass-dependent disk sizes and accretion rates. Since disks around low-mass stars are systematically smaller than those around higher-mass stars, dust grains are distributed within a more confined region closer to the central source, resulting in systematically higher temperatures. As a representative of the low-luminosity stars in the USco sample, we considered a model with $L_{\star}\,{=}\,0.1\,L_{\odot}$ (see the left panel of Figure~\ref{fig:sample}). Our relation predicts $T_{\rm dust}\,{=}\,24.8\,{\rm K}$. In contrast, the scalings of \citet{Andrews2013} and \citet{vanderPlas2016} result in temperatures of 14.1\,K and 15.2\,K, respectively, which are lower than our prediction by 10.7\,K and 9.6\,K.

To evaluate the role of viscous heating, we recalculated the 27 models in the $L_{\star}\,{=}\,0.1\,L_{\odot}$ bin without including viscous heating. The median difference in dust temperature between the accreting and non-accreting models is 5.2\,K, indicating that viscous heating contributes substantially to the higher temperatures predicted by our models. The remaining difference relative to the prescriptions of \citet{Andrews2013} and \citet{vanderPlas2016} arises from the updated disk structure adopted in this work, including the stellar-mass-dependent disk sizes. As discussed above, variations in the flaring index and scale height mainly contribute to the scatter in dust temperature at a given stellar luminosity.

\section{Statistics of the dust masses}
\label{sec:mdust}

To calculate the dust mass, we adopted the dust temperature predicted by the $T_{\rm dust}\,{-}\,L_{\star}$ relation given by Eq.~\ref{eqn:tdust} and the porous dust opacities $\kappa_{\rm 0.88mm}\,{=}\,0.84\,{\rm cm^2/g}$ described in Sect.~\ref{sec:dustopac}. For each source, we used the distance derived from Gaia parallaxes when converting the observed flux density to dust mass. If Gaia distances were unavailable, we adopted distances of 145\,pc and 160\,pc for the USco and Cha~I sources, respectively. These assumptions differ from those made in previous studies. 

\subsection{Cumulative distribution of the dust masses}

\begin{figure}
\centering
\includegraphics[width=0.8\textwidth]{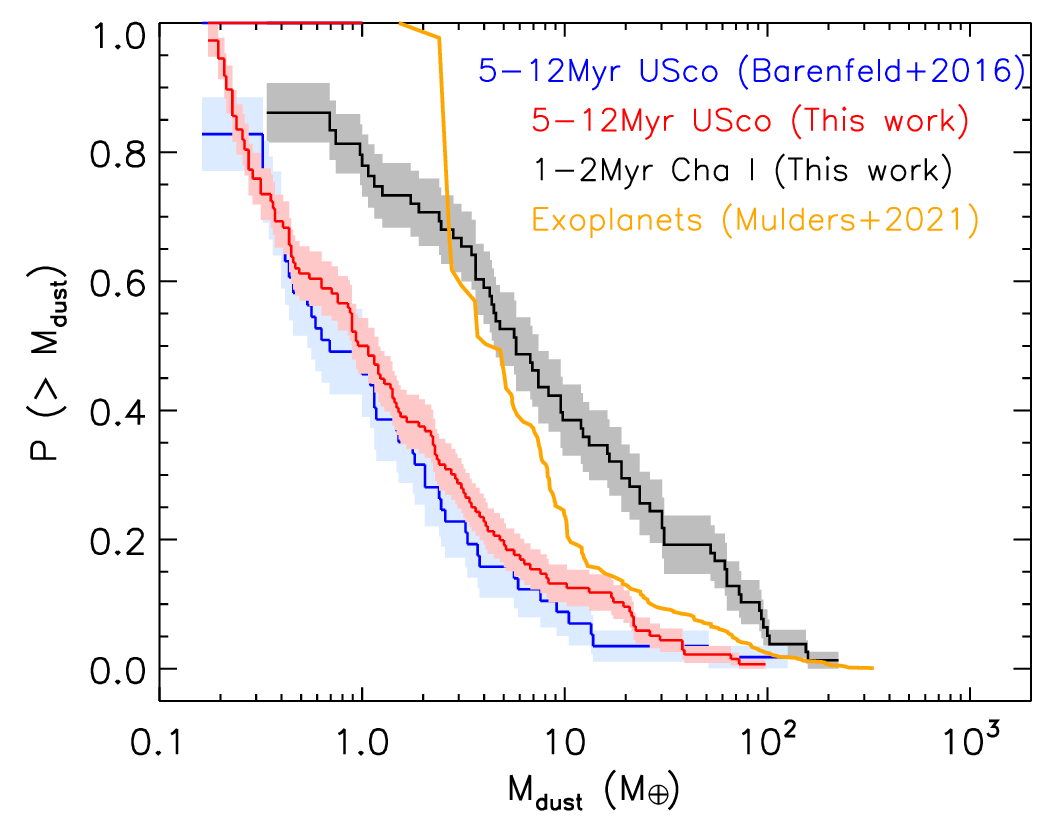}
\caption{Cumulative distributions of dust masses $M_{\rm dust}$. The black and red curves show the distributions for the young Cha~I and older USco disk populations, respectively. The blue curves represent the dust mass distribution derived by \citet{Barenfeld2016} for 57 full or transitional disks in USco. The orange curves show the distribution of exoplanet masses compiled by \citet{Mulders2021}.}
\label{fig:cdfmdust}
\end{figure}

Figure~\ref{fig:cdfmdust} presents the Kaplan–Meier estimator of the resulting dust mass distributions. For comparison, we also show the dust mass distribution derived by \citet{Barenfeld2016} for 57 full and transitional disks in USco. In their work, a uniform distance of 145\,pc was assumed for all sources, together with an absorption opacity of $\kappa_{\rm 0.88mm}\,{=}\,2.7\,{\rm cm^2\,g^{-1}}$ and the $T_{\rm dust}\,{-}\,L_{\star}$ scaling of \citet{Andrews2013}. Our inferred dust masses are slightly higher than those reported by \citet{Barenfeld2016}. The adoption of porous dust opacities increases the inferred dust masses by a factor of $2.7/0.84\,{=}\,3.2$. However, our revised  $T_{\rm dust}\,{-}\,L_{\star}$ relation predicts higher dust temperatures for low-mass stars. For a representative star with $L_{\star}\,{=}\,0.1\,L_{\odot}$, the Planck function at 0.88\,mm is larger by a factor of 2.34 than that obtained using the \citet{Andrews2013} temperature prescription, which reduces the inferred dust mass by the same factor. Consequently, the higher dust temperatures partially compensate for the lower opacity, resulting in a net increase in dust mass of a factor of only 1.37. In addition, the substantially larger sample analyzed in this work (136 sources) extends the dust mass distribution toward lower masses, providing a more complete census of the evolved disk population in USco. We also examined the optical-depth structure of the 918 radiative transfer models to assess the applicability of the optically thin approximation adopted in Eq.~\ref{eqn:mana}. For each model, we determined the radius at which the vertically integrated optical depth ($\tau_{\rm vert}$) at 0.88\,mm reaches unity. We find that the median value of this radius is only $0.3\,R_{\rm c}$, and in 74\% of the models the optically thick region is confined within $R_{\rm c}$. Furthermore, we also derived the fraction of dust mass that resides above the $\tau_{\rm vert}\,{=}\,1$ surface for each model. The median fraction is 82\%. These results indicate that, over the range of disk properties explored in our grid, the optically thick region is generally confined within the inner disk, while the majority of the disk mass remains in optically thin regions at millimeter wavelengths. This supports the applicability of the optically thin approximation adopted to estimate the dust masses in this work.

For consistency, we calculated the dust masses of the Cha~I disks using the same opacity, dust-temperature prescription, and updated distances. The dust masses of the USco disks are systematically lower than those of the younger Cha~I population, consistent with previous studies \citep{Pascucci2016,Manara2023}. The median dust mass of the USco sample is $0.95\,M_{\oplus}$, a factor of six lower than that of the Cha~I disks ($5.69\,M_{\oplus}$). Although the USco and Cha I samples span the same spectral-type range (K1–M5.5), their stellar-mass distributions are statistically different because the two-sample Kolmogorov–Smirnov test yields $p\,{<}\,0.01\%$. To assess whether this affects our comparison, we repeated the analysis using only sources with $0.08\,M_{\odot}\,{<}\,M_{\star}\,{<}\,0.13\,M_{\odot}$. We found median dust masses of $2.83\,M_{\oplus}$ for Cha~I and $0.44\,M_{\oplus}$ for USco, indicating that the younger Cha~I disks remain substantially more massive than USco disks. This result is consistent with the growing ALMA surveys demonstrating a systematic decline in millimeter dust masses with the age of star-forming regions \citep[e.g.,][]{Miotello2023}. Such a trend likely reflects the combined effects of grain growth into planetesimals, radial drift, and disk dispersal, which progressively reduce the reservoir of millimeter-sized particles responsible for the (sub-)millimeter continuum emission. Although the relative importance of these processes remains uncertain, the substantially lower dust masses observed in USco provide compelling evidence for significant evolution of the solid component over the ${\sim}\,5\,{-}\,10\,{\rm Myr}$ age interval separating the two regions.

A comparison between the masses of protoplanetary disks and mature exoplanetary systems has raised a potential ``mass budget problem'' for planet formation. Specifically, the dust masses inferred for many Class~II disks appear insufficient to account for the masses of known exoplanet populations, particularly giant planet systems \citep{Manara2018,Andrews2020}. Although observational selection effects partially alleviate this discrepancy \citep{Mulders2021}, the issue remains an active topic of debate \citep{Godines2026,Lee2026}. Several solutions have been proposed. One possibility is that planet formation begins during the earlier Class 0/I stages, when disks are more massive than their Class II counterparts \citep{Tychoniec2018,Tychoniec2020}. Alternatively, disks may be continuously replenished by material accreted from their natal environment, increasing the total mass available for planet formation over their lifetimes \citep{Manara2018,Gupta2024}. A third possibility is that (sub-)millimeter continuum observations systematically underestimate disk masses because the inner region and midplane of the disk likely remain optically thick and the adopted dust opacities are uncertain \citep{Ballering2019,Xin2023,Liuy2024}.

The orange curve in Figure~\ref{fig:cdfmdust} refers to the distribution of the exoplanet masses corrected for observational selection effects and detection biases \citep{Mulders2021}. The median mass of the exoplanets is $4.28\,M_{\oplus}$, slightly lower than the median dust mass of the Cha~I disks ($5.69\,M_{\oplus}$). We note that our analysis is restricted to host stars with spectral types between K1 and M5.5 (Sect.~\ref{sec:sample}). Including earlier-type stars will shift the dust mass distribution to higher values. In contrast, the median mass of the exoplanets is more than four times higher than the median dust mass of the USco disks ($0.95\,M_{\oplus}$). This comparison has important implications for the mass-budget problem of planet formation. While the dust reservoirs in the young Cha~I disks appear broadly consistent with the masses of mature planetary systems, the significantly lower dust masses observed in USco suggest that a large fraction of the solid material has already been removed from the observable millimeter-sized dust population by ages of ${\sim}\,5\,{-}\,12\,{\rm Myr}$. The missing solids may have been incorporated into larger bodies such as planetesimals and planets, lost through radial drift, or hidden in optically thick regions of the disk. Consequently, the low dust masses of the USco disks do not necessarily imply a lack of material for planet formation, but instead provide evidence that substantial processing of the primordial dust reservoir has already taken place by this evolutionary stage.

\subsection{Correlation between dust masses and stellar masses}

\begin{figure*}
\centering
\includegraphics[width=0.8\textwidth]{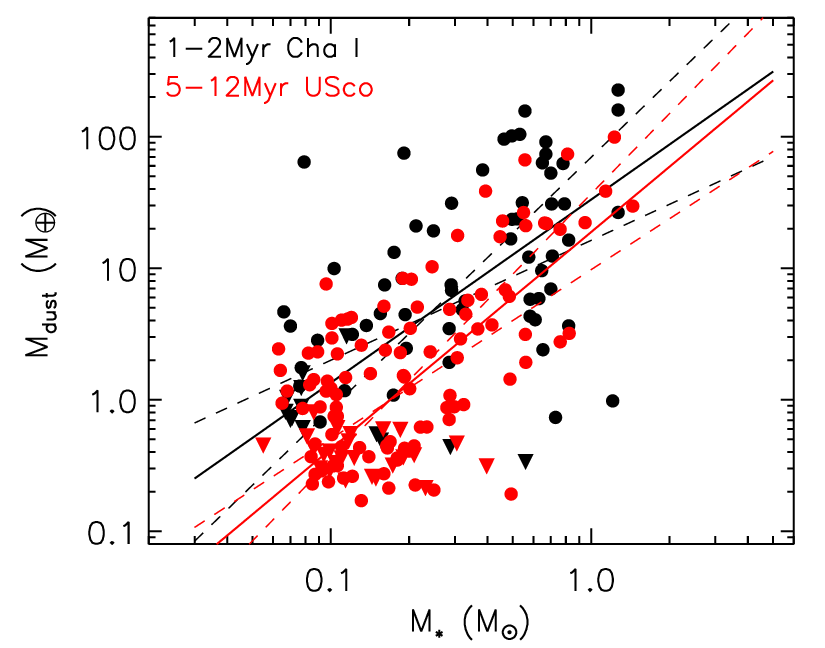}
\caption{dust mass as a function of stellar mass. Black symbols show the results for Cha~I disks, while red symbols represent the USco sources. Millimeter detections are indicated with dots, whereas triangles mean upper limits of the millimeter flux are reported. The solid lines are the best-fit $M_{\star}\,{-}\,M_{\rm dust}$ relation, while the dashed lines mark the 95\% confidence intervals on the relation.}
\label{fig:mstarmdust}
\end{figure*}

Previous studies have shown that disk dust mass scales with stellar mass \citep[e.g.,][]{Andrews2013,Ansdell2016}. To assess the significance of this relation, we analyzed the USco and Cha~I samples separately using the \texttt{ASURV} package, which properly accounts for censored data. For both regions, a Cox proportional hazards test rejects the null hypothesis of no correlation with $p\,{<}\,10^{-4}$. This result is independently confirmed by Spearman's rank correlation tests, which likewise yield $p\,{<}\,10^{-4}$ for both samples. These statistical tests demonstrate a highly significant correlation between stellar mass and dust mass in both USco and Cha~I. We then quantified the $M_{\star}\,{-}\,M_{\rm dust}$ relation using the Bayesian linear regression routine \texttt{Linmix\_err}, which accounts for measurement uncertainties and upper limits. The best relations are
\begin{equation}
 {\rm log }\,M_{\rm dust}\,{=}\,1.28_{-0.14}^{+0.15}\,{+}\,1.65_{-0.19}^{+0.19}\,{\rm log}\,M_{\star}
\end{equation}
for the USco disks and 
\begin{equation}
 {\rm log }\,M_{\rm dust}\,{=}\,1.52_{-0.15}^{+0.16}\,{+}\,1.38_{-0.24}^{+0.26}\,{\rm log}\,M_{\star}
\end{equation} 
for the Cha~I sources. The best-fit relations and their 95\% confidence intervals are shown in Figure~\ref{fig:mstarmdust}. For comparison, \citet{Pascucci2016} derived ${\rm log }\,M_{\rm dust}\,{=}\,0.8(\pm 0.2)\,{+}\,1.9(\pm 0.4)\,{\rm log}\,M_{\star}$ for USco and ${\rm log }\,M_{\rm dust}\,{=}\,1.1(\pm 0.1)\,{+}\,1.3(\pm 0.2)\,{\rm log}\,M_{\star}$ for Cha~I. The slopes derived in this work are consistent with those reported by \citet{Pascucci2016} within the quoted $1\sigma$ uncertainties, indicating that the updated analysis does not produce a statistically significant change in the stellar-mass dependence of the dust masses. The small differences in the best-fit parameters likely arise from several improvements adopted in this work, including the updated ALMA flux densities for 14 Cha~I disks presented by \citet{Long2018}, the substantially expanded USco sample compiled by \citet{Carpenter2025}, and the revised dust masses derived using Gaia distances together with the new $T_{\rm dust}$--$L_{\star}$ relation.

Demographic surveys of protoplanetary disks in different star-forming regions found that the $M_{\rm dust}\,{-}\,M_{\star}$ relation steepens with
time \citep[e.g.,][]{Pascucci2016,Ansdell2017}. The slightly steeper relation observed in the old USco population compared with Cha~I is broadly consistent with this evolutionary trend. Dust evolution models indicate that such behavior can arise because millimeter-sized grains are depleted more rapidly in disks around low-mass stars through the combined effects of grain growth, fragmentation, and radial drift \citep{Pascucci2016,Pinilla2022}. However, the evolution of the relation may be modified by dust trapping in pressure maxima, which can efficiently retain millimeter-sized particles and slow their inward drift \citep{Pinilla2020}. This highlights the important role of disk substructures in shaping the observed $M_{\rm dust}\,{-}\,M_{\star}$ relation. If pressure bumps are the primary mechanism regulating the retention of millimeter-sized pebbles, the observed steepening of the $M_{\rm dust}\,{-}\,M_{\star}$
relation with age implies that dust trapping efficiency likely depends on stellar mass. In particular, the preferential depletion of dust around lower-mass stars would suggest that pressure bumps are less effective at retaining pebbles in these disks. Such a stellar-mass dependence could arise if the properties of pressure bumps, such as their amplitude, lifetime, or spatial distribution, vary systematically with disk mass or stellar mass. Future theoretical and observational studies are needed to investigate this possibility.

\section{Summary}
\label{sec:summary}
We derived dust masses of 136 full and transitional disks in the Upper Scorpius association using the recently expanded ALMA 0.88\,mm continuum sample compiled by \citet{Carpenter2025}. Our analysis incorporates a new calibration of the $T_{\rm dust}\,{-}\,L_{\star}$ relation based on self-consistent radiative transfer models that account for stellar-mass-dependent disk sizes, accretion rates, and porous dust properties. The main results are summarized as follows.

\begin{itemize}
\item[(1)] We constructed a grid of 918 radiative transfer models spanning stellar luminosities from $10^{-3}$ to $100\,L_{\odot}$. The resulting mass-averaged dust temperatures are well described by ${\rm log}\,T_{\rm dust}=1.506+0.115\,{\rm log}\,L_{\star}+0.004\,({\rm log}\,L_{\star})^2$. Compared with previous prescriptions, our relation predicts systematically higher dust temperatures in the low-stellar-mass regime, largely due to the fact that disks around low-mass stars are observationally found to be smaller and therefore warmer.

\item[(2)] Using the revised dust temperatures together with porous dust opacities ($\kappa_{\rm 0.88mm}\,{=}\,0.84\,{\rm cm^2\,g^{-1}}$), we derived dust masses for the USco sample. The median dust mass of the USco disks ($0.95\,M_{\oplus}$) is approximately six times lower than that of a comparison sample of young Cha~I disks ($5.69\,M_{\oplus}$). This result provides further evidence that the reservoir of millimeter-sized dust grains declines substantially over the first few Myr of disk evolution.

\item[(3)] The median dust mass of the USco disks is also substantially lower than the median mass of mature exoplanetary systems. This comparison suggests that by ages of ${\sim}\,5{-}12\,{\rm Myr}$, a large fraction of the primordial solid reservoir has either been incorporated into larger bodies, removed through radial drift, or become hidden from millimeter observations.

\item[(4)] We confirm a highly significant correlation between dust mass and stellar mass in both USco and Cha~I. The best-fit relations are
${\rm log}\,M_{\rm dust}\,{=}\,1.28\,{+}\,1.65\,{\rm log}\,M_{\star}$ for USco and ${\rm log}\,M_{\rm dust}\,{=}\,1.52\,{+}\,1.38\,{\rm log}\,M_{\star}$ for Cha~I. The slightly steeper relation in the older USco population is consistent with previous observational studies and with dust evolution models in which millimeter-sized grains are depleted more efficiently around low-mass stars.
\end{itemize}

\section*{Acknowledgements}
YL acknowledges financial supports by the Natural Science Foundation of Sichuan Province of China (grant no. 2025ZNSFSC0060), the Fundamental Research Funds for the Central Universities (grant no. 2682025CX028), and the International Partnership Program of Chinese Academy of Sciences (grant no. 019GJHZ2023016FN). MF acknowledges the financial support by the National Natural Science Foundation of China (Grant no. 12573028). This work is supported by the China Manned Space Program with grant nos. CMS-CSST-2025-A15 and CMS-CSST-2025-A16. We thank John M. Carpenter for insightful discussions.

\bibliographystyle{raa}
\bibliography{ms2026-0395}

\end{document}